\documentclass[12pt,preprint]{aastex}
\usepackage{multirow}
 \newcommand{\beq}{\begin{equation}}
 \newcommand{\eeq}{\end{equation}}
 \newcommand{\beqn}{\begin{eqnarray}}
 \newcommand{\eeqn}{\end{eqnarray}}
 \newcommand{\non}{\nonumber}
 \newcommand{\no}{\noindent}

\newcommand{\pa}{\partial}

\begin{document}
\title{Galaxies with Supermassive Binary Black Holes:\\
(III) The Roche Lobes and Jiang-Yeh Lobe in a Core System}

\author{Li-Chin Yeh$^{1}$ and Ing-Guey Jiang$^{2}$}

\affil{
{$^{1}$Department of Applied Mathematics,}\\
{ National Hsinchu University of Education, Hsin-Chu, Taiwan}\\ 
{$^{2}$Department of Physics and Institute of Astronomy,}\\
{ National Tsing-Hua University, Hsin-Chu, Taiwan}
}

\email{jiang@phys.nthu.edu.tw}

\begin{abstract} 
Three-dimensional equi-potential surfaces of a galactic 
system with supermassive binary black holes are discussed herein.
The conditions of topological transitions for the important surfaces,
i.e. Roche Lobes and Jiang-Yeh Lobe,
are studied in this paper. In addition, the mathematical properties of 
the Jacobi surfaces   
are investigated analytically.
Finally, a numerical procedure for determining the regions of the 
Roche Lobes and Jiang-Yeh Lobe is suggested. 
\end{abstract}

\newpage
\section{Introduction}

It is known that most galaxies, if not all, host  
supermassive black holes, as implied by the
stellar kinematic data and the radiation 
of active galactic nuclei near the centers of galaxies. 
Among these, early-type galaxies were  likely  constructed through
the merging events of spiral galaxies. Therefore, they might host more than one
supermassive black holes. The evolution of these 
supermassive black holes shall be related to the merging processes and 
the properties of those involved galaxies.
Supermassive black holes will then scatter their surrounding stars and 
therefore influence the structures of galaxies as feedback.
In fact, the possible co-evolution of galaxies and supermassive black holes 
has become one of the most important and popular astronomical 
topics in recent years. 

In a merged galactic system, supermassive black holes  
will be driven to the galactic center by dynamical friction.
When there is more than one supermassive black holes, i.e. two or three, 
it is likely that supermassive binary black holes (SBBH) will form in the 
galactic center. A configuration of three or more such bodies is less stable
due to possible dynamical instabilities among three or multiple bodies. 
In order to study the evolution of SBBH,
Quinlan (1996) carried out  numerical experiments on the stellar scattering
by SBBH in the form of  restricted three-body problems.
Indeed, the restricted three-body problem has been a useful model 
in many fields,
such as the evolution of binary stars (Eggleton 1983)  
and orbital calculations of planetary systems 
(Mikkola et al. 1994; Namouni 1999). 
Moreover, modifications of the restricted three-body
problem were employed to model various problems related to
planetary systems
(Chermnykh 1987; Papadakis 2004, 2005a, 2005b;
Jiang \& Yeh 2006; Yeh \& Jiang 2006;
Perdios, Kalantonis \& Douskos 2008;
Kushvah 2008a, 2008b, 2009, 2011a, 2011b; Douskos 2011;
Kushvah, Kishor \& Dolas 2012; Kishor \& Kushvah 2013; 
Douskos et al. 2012; Douskos 2015).

On the other hand, n-body simulations also play important roles
in the study of galactic mergers (Wu \& Jiang 2009, 2012, 2015). 
Milosavljevic \& Merritt (2001) employed n-body simulations to study
the orbital decay of black holes and
the formation of SBBHs during galactic mergers.
Once an SBBH is formed at the galactic center, it is important to 
understand how it affects the central surface brightness of a galaxy.
If that can be done, the relationship between the
surface brightness profiles and properties of SBBH 
might be established. 

Motivated by that goal of establishing 
SBBH--brightness connections, Jiang \& Yeh (2014a) took a first step
to construct a theoretical model with an SBBH embedded in 
a galactic center. Their employed galactic density profile was
the same as one of those suggested in Kandrup et al. (2003), i.e.:
\beq
\rho(\bar{r})=\rho_c\left(\frac{\bar{r}}{r_b}\right)^{-\gamma}
\left\{1+\left(\frac{\bar{r}}
{r_b}\right)^{\alpha}\right\}^{\frac{\gamma-\beta}{\alpha}}\,,
\label{rho_bar_r1}
\eeq
where $\rho_c$, $r_b$, $\alpha$, $\beta$ and $\gamma$ are the constants.
In fact, these density profiles are equivalent to the 
$(\alpha, \beta, \gamma)$-models in Hernquist (1990), albeit with 
different parameter definitions; it is a five-parameter generalization 
of Jaffe's double-power law model.
Zhao (1996) found that the gravitational potentials corresponding to 
these density profiles could have simple analytic forms for certain choices
of $\alpha$, $\beta$ and $\gamma$. Jiang \& Yeh (2014a) investigated the 
existence of equilibrium points, i.e. 
Lagrange Points and Jiang-Yeh Points,
for a system with $\alpha=2$, $\beta=4$ and $\gamma=0$,
i.e. a core density profile. 
They proved that Jiang-Yeh Points exist if, and only if,  
the galactic mass is larger than a critical mass. 
Moreover, Jiang \& Yeh (2014b) found that 
Jiang-Yeh Points always exist for a system with 
$\alpha=2$, $\beta=5$ and  $\gamma=1$, i.e. a cuspy density profile.

In these models, the orbital motion of stars near the SBBH at the 
centers of galaxies can be determined conveniently 
as a modification of restricted three-body systems.
Because the time-scales of the SBBH orbital changes and
the variations of overall galactic potential are much larger than the
dynamical time near the galactic center, the stellar orbits 
derived from these models are good approximations 
of realistic n-body simulations.
The surface brightness of galactic centers can be calculated when 
stellar orbits have been obtained.
The surface brightness near the center of galaxies might 
show some signals which are related to 
the dynamical evolution of stars near the SBBH.  
Therefore, these models can become
useful tools for interpreting the properties of SBBH 
from the observational surface brightness of galaxies.

For such an SBBH model, it would be interesting to understand precisely
how stars move around supermassive black holes. It is important to 
investigate whether stars would 
stay within an equi-potential surface centered on one 
supermassive black hole, or jump to another one 
under particular conditions.  
For the above purpose, we will present the shape and topology 
of equi-potential surfaces, and be able to determine whether stars are
inside or outside one of equi-potential surfaces in three 
dimensional space.   

Thus, the main goal of this paper is to extend the model in 
Jiang \& Yeh (2014a)
into a three-dimension model and present the equi-potential surface.
The condition under which the important surfaces would have 
topological changes will also be investigated. 

Moreover, we will also analytically study the properties of 
the Jacobi integral of equi-potential surfaces, 
and suggest a numerical procedure 
to determine the regions of
important equi-potential surfaces.
  
The model is given in Section 2,
the equi-potential surfaces are presented in Section 3,
the mathematical properties of the Jacobi surfaces
are studied in Section 4, the determination of 
regions of Roche Lobes and Jiang-Yeh Lobe is in Section 5,
and  the concluding remarks are presented in Section 6.

\section{The Model}

The dynamics of a star near the center of a galaxy
in a three-dimensional space is studied herein.
The star is governed by 
the galactic potential and the gravitational force from 
the central SBBH. As in Jiang \& Yeh (2014a),   
the star is treated as a test particle, 
the galactic potential is fixed, and the SBBH is an equal-mass pair moving
along a circular orbit. The galactic density distribution could have any form 
as long as it is spherically symmetric. 
A power law with a core, as the profile used in Jiang \& Yeh (2014a),
is adopted herein: 
\beq
\rho(\bar{r})=\rho_c 
\left\{1+\left(\frac{\bar{r}}
{r_b}\right)^{2}\right\}^{-2}\,,
\label{rho_bar_r}
\eeq
where $\rho_c$ is a constant.
The scale length $r_b$ is called the break radius.
When $\bar{r} \gg r_b$, the above becomes a power law with index
$4$. For $\bar{r} \ll r_b$, the density approaches becoming a constant 
as $\bar{r}$ gets close to zero. 
Thus, the break radius, $r_b$,  marks a transition from the 
outer power law with index $4$
to the core of the system. It is also called the core radius here. 
The mass of the galaxy up to a radius $\bar{r}$ is expressed as :
\beq
M(\bar{r})= 4\pi \int^{\bar{r}}_0 r_{du}^2\rho(r_{du})dr_{du}
=2\pi\rho_c\left\{\tan^{-1}(\bar{r}/r_b)
-\frac{(\bar{r}/r_b)}{1+(\bar{r}/r_b)^2}\right\},
\eeq
where $r_{du}$ is a dummy variable of the integration.
Note that the galactic mass $M(\bar{r})$ does not include the mass of SBBH.
The galactic total mass is set to be $M_g$; thus 
the constant $\rho_c={M_g}\pi^{-2}$.

Moreover, 
the corresponding potential is expressed as:
\beq
V(\bar{r})= -4\pi\left[\frac{1}{\bar{r}}
\int_0^{\bar r}\rho(r_{du}) r_{du}^2 d r_{du} 
+\int_{\bar r}^\infty \rho (r_{du})r_{du} dr_{du}\right]
=
-c\frac{\tan^{-1}(\bar{r}/r_b)}{(\bar{r}/r_b)},\label{eq:va1}
\eeq
where $c=2M_{g}\pi^{-1}$.
With this galactic potential
and the SBBH at $(-\bar{R}, 0,0)$, $(\bar{R}, 0,0)$, 
the equations of motion of a star 
under the rotating frame in a three-dimensional space 
are now written as
\beq \left\{
\begin{array}{ll}
&  \frac{d\bar{x}}{d\bar{t}}=\bar{u}  \\
& \frac{d\bar{y}}{d\bar{t}}=\bar{v}  \\
& \frac{d\bar{z}}{d\bar{t}}=\bar{w} \\
& \frac{d\bar{u}}{d\bar{t}}=2\bar{n}\bar{v} +\bar{n}^2\bar{x}-\frac{G\bar{m}(
\bar{x}+\bar{R})}{\bar{r}_1^3}-\frac{G\bar{m}(\bar{x}-\bar{R})}
{\bar{r}_2^3}-\frac{\pa V}{\pa \bar{r}}\frac{d \bar{r}}{d\bar{x}}\\
& \frac{d\bar{v}}{d\bar{t}}=-2\bar{n}\bar{u}+\bar{n}^2 \bar{y}-\frac{G\bar{m}\bar{y}}{\bar{r}_1^3}-\frac{G\bar{m}\bar{y}}{\bar{r}_2^3}
-\frac{\pa V}{\pa \bar{r}}\frac{d \bar{r}}{d\bar{y}} \\
&\frac{d\bar{w}}{d\bar{t}}=-\frac{G\bar{m}\bar{z}}{\bar{r}_1^3}-\frac{G\bar{m}\bar{z}}{\bar{r}_2^3}
-\frac{\pa V}{\pa \bar{r}}\frac{d \bar{r}}{d\bar{z}},
\end{array}  \right. \label{eq:3body2}
\eeq
where $\bar{n}$ is the SBBH's angular velocity,
$$ \bar{r}=\sqrt{\bar{x}^2+\bar{y}^2+\bar{z}^2},
\bar{r}_1=\sqrt{(\bar{x}+\bar{R})^2+\bar{y}^2+\bar{z}^2}, \quad {\rm and}\,
\quad \bar{r}_2=\sqrt{(\bar{x}-\bar{R})^2+\bar{y}^2+\bar{z}^2}. $$
As in Yeh, Chen, \& Jiang (2012) and Jiang \& Yeh (2014a), 
we can employ non-dimensionalization to simplify the above equations.
When  non-dimensional variables are set as
$$x=\frac{\bar{x}}{L_0}, y=\frac{\bar{y}}{L_0},z=\frac{\bar{z}}{L_0},
r=\frac{\bar{r}}{L_0}, R=\frac{\bar{R}}{L_0}, \,\,
 t=\frac{\bar{t}}{t_0},  m=\frac{\bar{m}}{m_0}, u=\frac{\bar{u}}{u_0},
v=\frac{\bar{v}}{u_0}, w=\frac{\bar{w}}{u_0}, n={t_0}{\bar{n}},
$$
and some of the above parameters are further assumed to satisfy  
$u_0=\frac{L_0}{t_0}$,  $Gm_0=\frac{L_0^3}{t_0^2}$, $L_0=r_b$,
System (\ref{eq:3body2}) can be rewritten as
\beq \left\{
\begin{array}{ll}
&  \frac{dx}{dt}=u  \\
& \frac{dy}{dt}=v  \\
& \frac{dz}{dt}=w  \\
& \frac{du}{dt}=2nv +n^2 x-\frac{m(x+R)}{r_1^3}-\frac{m(x-R)}
{r_2^3}+\frac{c x}{r} \left\{\frac{1}{(1+r^2)r}-\frac{\tan^{-1} r}{r^2}
\right\} \\
& \frac{d v}{dt}=-2nu+n^2 y-\frac{my}{r_1^3}-
\frac{my}{r_2^3}+\frac{c y}{r}
 \left\{\frac{1}{(1+r^2)r}-\frac{\tan^{-1} r}{r^2}\right\}\\
& \frac{d w}{dt}=-\frac{mz}{r_1^3}-\frac{mz}{r_2^3}+\frac{c z}{r}
 \left\{\frac{1}{(1+r^2)r}-\frac{\tan^{-1} r}{r^2}\right\}.
\end{array}  \right. \label{eq:3body3_non}
\eeq
Moreover, in the above equations, 
under the influence of the galactic potential, 
the mean motion, i.e.
angular velocity, of each component of SBBH satisfies
\beq
m n^2 R = \frac{m^2}{(2R)^2}+ m |f_g(R)|.
\label{eq:n2}
\eeq
That is, the centrifugal force of a black hole is equal to 
the total of the forces from another black hole and 
the background galaxy.      
The gravitational force per unit mass at $r$ 
from the galactic potential, i.e. $f_g(r)$,  can be expressed as 
\beq
f_g(r)
= c \left\{\frac{1}{(1+r^2)r}-\frac{\tan^{-1} r}{r^2}
\right\}.
\label{eq:fg}
\eeq
Therefore,
\beq
n= \left\{ \frac{m}{4R^3}+\frac{1}{R}|f_g(R)|\right\}^{1/2}.
\label{eq:n2}
\eeq
The corresponding Jacobi integral of this system is 
\beq
C_J = 2U-\dot{x}^2 -\dot{y}^2 -\dot{z}^2= n^2 (x^2+y^2)
       + \frac {2m}{r_1} + \frac {2m}{r_2}
       + 2c\frac{ {\rm tan}^{-1}(r) }{r}-\dot{x}^2-\dot{y}^2 -\dot{z}^2,
\label{eq:Ja1}
\eeq
\no where $r=\sqrt{x^2+y^2+z^2}$, $r_1=\sqrt{(x+R)^2+y^2+z^2}$, and
$r_2=\sqrt{(x-R)^2+y^2+z^2}$.

\section{Equi-Potential Surfaces: Roche Lobes and Jiang-Yeh Lobe}

As shown in last section, we have extended our equations of motion from 
two-dimensional space (Jiang \& Yeh 2014a) to
three-dimensional space.
Because the SBBH is moving on the $xy$-plane, 
all results about equilibrium points presented in Jiang \& Yeh (2014a)  
are also valid here. Thus,
Jiang-Yeh Points could exist and
would influence the structures of 
equi-potential surfaces in the system. Those equi-potential surfaces
cutting through equilibrium points would have the shapes of lobes. 
Two lobes centering on black holes are still called 
Roche Lobes, and the new one cutting through the Jiang-Yeh Points are called
Jiang-Yeh Lobe in this paper.

Roche Lobes of the standard restricted three-body problem 
are equi-potential surfaces cutting through the first Lagrange Point
L1. Each lobe centers on one star and connects with another lobe
through L1. 
On the rotating frame of this binary system,
test particles are locally force-free along 
the directions which are tangential to the equi-potential surfaces.
This is why the Roche-lobe-over-flow could accomplish the 
mass transfer from one star to another (see Fig. 1(a)). 

The Jacobi integral in Eq. (\ref{eq:Ja1})
contains the potential part and the kinetic part.
Taking $\dot{x}=\dot{y} =\dot{z}=0$ in Jacobi integral
would lead to equi-potential surfaces (equivalently, 
zero-velocity surfaces).
For convenience, this particular case of the Jacobi integral
is set as 
\beq
F(x,y,z)\equiv n^2 (x^2+y^2) + \frac {2m}{r_1} + \frac {2m}{r_2}
       + 2c\frac{ {\rm tan}^{-1}(r) }{r},\label{eq:f1}
\eeq
and $F(x,y,z)$ is called {\it the Jacobi surfaces} here. 
For a given value of $J$,
$F(x,y,z)=J$ is an equi-potential surface.
The equi-potential surface cutting through a particular equilibrium 
point can be easily determined. 
For example,
the coordinate of a given equilibrium point  
$(x_E, y_E, z_E)$ is inserted into
$F(x,y,z)$, and
we set the constant $F(x_E, y_E, z_E) \equiv J_{E}$.
Thus, $F(x,y,z)=J_{E}$ is an equi-potential 
surface cutting through this equilibrium point.
Moreover, $F(x,y,0)=J_{E}$ defines an equi-potential curve
(zero-velocity curve) on the $xy$-plane,
and $F(x,0,z)=J_{E}$ defines an equi-potential curve
on the $xz$-plane. 

The convention of the names of equilibrium points here 
is the same as in Jiang \& Yeh (2014a). That is,  
the equilibrium point at the origin (0,0,0) is Lagrange Point 1 (L1), 
the one at $x$-axis with $x>R$ is Lagrange Point 2 (L2),  
the one at $x$-axis with $x<-R$ is Lagrange Point 3 (L3), 
the one at $y$-axis with $y>0$ is Lagrange Point 4 (L4), 
the one at $y$-axis with $y<0$ is Lagrange Point 5 (L5), 
the one at $x$-axis with $0<x<R$ is Jiang-Yeh Point 1 (JY1), 
the one at $x$-axis with $-R<x<0$ is Jiang-Yeh Point 2 (JY2).

In order to present these equi-potential surfaces and curves,
Models A, B, C and D are chosen as four cases with different 
sets of parameters. The values of their parameters are listed in 
the top three rows of Table 1. The resulting plots are shown
in Figs.1-4, respectively. Because we consider equal mass SBBH, 
the system is symmetric. All important points we define and mark 
in Figs. 1-4 are in the regions with non-negative $x$, $y$, $z$
coordinates.

In Model A, our system's parameters are $M_g=0$, $m=R=1$.
The L1-passing equi-potential surfaces, i.e. the 
Roche Lobes, are shown in Fig. 1(a).
Both the L1-passing and L2-passing equi-potential curves on 
the $xy$-plane and $xz$-plane are presented in Figs. 1(b)-(c).
On the $yz$-plane, the equi-potential curves with $x=R$,
i.e. $F(R,y,z)=J_{L1}$ and  $F(R,y,z)=J_{L2}$, are plotted
in Fig. 1(d). From these plots, it is clear that two Roche Lobes
are connected at L1, which is simply the same as 
in the classic binary models.
In Figs. 1(b)-(c), the equi-potential curve $F(x,y,0)=J_{L1}$ 
or $F(x,0,z)=J_{L1}$ goes through
$x-$axis at L1, $(0,0, x_{L1})$, 
and another point which is set as $(x_a,0,0)$.

The intersection between
the equi-potential curve $F(x,y,0)=J_{L1}$ and the line $x=R$
is set as $(R,y_a,0)$; the intersection between
the equi-potential curve $F(x,y,0)=J_{L2}$ and the line $x=R$
is set as  $(R,y_b,0)$; the intersection between
the equi-potential curve $F(x,0,z)=J_{L1}$ and the line $x=R$
is set as  $(R,0,z_a)$; the intersection between
the equi-potential curve $F(x,0,z)=J_{L2}$ and the line $x=R$
is set as  $(R,0,z_b)$.

In Model B, our system's parameters are set as $M_g=10$, $m=R=1$.
The structures of the equi-potential surfaces and curves are 
presented in Fig. 2. 
From Figs. 2(b)-(c), we find that the L1-passing equi-potential curves
are not closed, while the L2-passing equi-potential curves
form closed loops. Thus, the L2-passing equi-potential
surfaces in three dimensional space are defined as the Roche Lobes, as
shown in Fig. 2(a). The equi-potential curves on the $yz$-plane 
with $x=R$, i.e. $F(R,y,z)=J_{L1}$ and $F(R,y,z)=J_{L2}$, are plotted
in Fig. 2(d). 
In Figs. 2(b)-(c), the equi-potential curve $F(x,y,0)=J_{L2}$ 
or $F(x,0,z)=J_{L2}$ 
goes through $x-$axis at L2, $(0,0, x_{L2})$, and another point 
which is set as $(x_b,0,0)$.
The marked $y_a$, $y_b$, $z_a$ and $z_b$ were defined previously, as in Fig. 1.
              
In Model C, when our system's parameters are $M_g=30$, $m=R=1$,
Jiang-Yeh Points, i.e. JY1 and JY2, exist. 
As shown in Figs. 3(b)-(c), 
the equi-potential curves which pass JY1 and JY2 
form a central closed loop. The L2-passing equi-potential curves
also form closed loops centered on black holes.
The JY1-JY2-passing equi-potential surfaces in three dimensional 
space are defined as Jiang-Yeh Lobe, and the L2-passing equi-potential
surfaces in three dimensional space are still 
defined as the Roche Lobes here, as presented in Fig. 3(a). 
The equi-potential curve on the $yz$-plane 
with $x=0$, i.e. $F(0,y,z)=J_{JY1}$, 
is plotted in Fig. 3(d).
The marked $x_b$, $y_b$ and $z_b$ were defined previously as in Figs. 2 and 1.  
The coordinate of JY1 is $(0,0, x_{JY1})$.   
The intersection
between $F(x,y,0)=J_{JY1}$ and the line $x=R$
is set as  $(R,y_c,0)$; the intersection
between $F(x,y,0)=J_{JY1}$ and the line $x=0$, i.e. $y$-axis,
is set as  $(0,y_d,0)$.
The intersection
between $F(x,0,z)=J_{JY1}$ and the line $x=R$
is set as  $(R,0,z_c)$; the intersection
between $F(x,0,z)=J_{JY1}$ and the line $x=0$
is set as  $(0,0,z_d)$.

In Model D, when the parameters are set as $M_g=30$, $m=1$, $R=2$,
Jiang-Yeh Points also exist. 
In Figs. 4(b)-(c), the JY1-JY2-passing equi-potential 
curves form three connected closed loops.
Thus, in three dimensional space, the JY1-JY2-passing equal potential 
surfaces form three lobes, as presented in Fig. 4(a).  
The central one is defined as the Jiang-Yeh Lobe, and the outer two are
the Roche Lobes.
The equi-potential curve $F(x,y,0)=J_{JY1}$ or $F(x,0,z)=J_{JY1}$ goes through
$x-$axis at JY1 and another point which is set as $(x_c,0,0)$.
The marked $y_b$, $y_c$, $y_d$, $z_b$, $z_c$ and $z_d$ were defined previously.
The equi-potential curves on the $yz$-plane 
with $x=0$, i.e. $F(0,y,z)=J_{L1}$ and $F(0,y,z)=J_{L2}$, are plotted
in Fig. 4(d). 

Apparently, the existence of galactic mass $M_g$ changes the system's
dynamical structure and separates two Roche Lobes. 
When $m=R=1$ and $M_g$ is between 
0 and 10, i.e. Models A and B, there are no Jiang-Yeh Points,
but the Roche Lobes are already separated because, as shown in Figs. 2(b)-(c),
the L1-passing equi-potential curves, $F(x,y,0)=J_{L1}$ 
and $F(x,0,z)=J_{L1}$, do not turn back to go across the $x$-axis.
Thus, the L2-passing equi-potential surfaces become the ones to form
Roche Lobes.  

Comparing Fig. 2(b) with Fig. 1(b), it is realized 
that there must be a transition
with a critical 
mass $M_g$ such that $J_{L1}=J_{L2}$, i.e. 
$F(x_{L1},y_{L1},z_{L1})=F(x_{L2},y_{L2},z_{L2})$. 
In order to search for this critical value of $M_g$ for any given $m$, $R$, 
we define a function
\beqn
T_1(m,R,M_g)
&\equiv&F(x_{L1},y_{L1},z_{L1})-F(x_{L2},y_{L2},z_{L2})\non \\
&=&F(0,0,0)-F(x_{L2},0,0)\non \\
&=&\frac{4m}{R}-n^2x^2_{L2}
-\frac{2m}{x_{L2}+R}-\frac{2m}{x_{L2}-R}+\frac{4M_g}{\pi}\left(1-
\frac{\tan^{-1}x_{L2}}{x_{L2}}\right). \label{eq:G0}
\eeqn
If there is a zero point of function $T_1$ for a set of parameters 
$m$,$R$ and $M_g$, this transition exists.
From Eq.(\ref{eq:G0}), the critical mass, denoted as $M_1$, 
can be expressed as Eq. (\ref{eq:mc_ast}):
\beq
M_1=\frac{\pi\left[n^2x_{L2}^2+\frac{2m}{x_{L2}+R}+\frac{2m}{x_{L2}-R}-
\frac{4m}{R}\right]}{4\left(1-\frac{\tan^{-1}x_{L2}}{x_{L2}}\right)}.
\label{eq:mc_ast}
\eeq
We numerically solve $M_1$ as a function of $R$ for different values of $m$, 
as presented in Fig. 5(a).  
For the case with $m=R=1$, we find that $M_1 \sim 2.11$. 
If the critical mass $M_1$ exists,
whether the total galactic mass $M_g$ is larger than $M_1$
would determine the topology of the Roche Lobes.
When $M_g<M_1$,
two Roche Lobes are connected as the case shown
in Fig. 1. When $M_g>M_1$, two Roche Lobes 
would be separated as shown in Figs. 2 and 3.

The difference between Figs.2 and 3 is whether or not there are 
Jiang-Yeh Points. There is a critical mass $M_c$ to determine
the existence of Jiang-Yeh point (Jiang \& Yeh 2014a) and
$M_c\sim 17.51$ when $m=R=1$.
Two Roche Lobes are separated in both plots 
due to $M_g > M_1$ for both Figs. 2 and 3.
In Fig. 3, $M_g > M_c$, so there are Jiang-Yeh Points, and 
thus a Jiang-Yeh Lobe exists and is  
located between the separated Roche Lobes.

When $m=1$ and $R=2$, we 
find that $T_1(1,2,M_g)>0$ for any positive $M_g$. In this case,
there is no zero point of $T_1$ and thus no critical mass $M_1$.
When no critical mass $M_1$ can be found, 
there is no such transition here.
  
On the other hand, 
it is shown that Roche Lobes are disconnected from the  
Jiang-Yeh Lobe in Fig. 3, but connected with the Jiang-Yeh Lobe
in Fig. 4. 
Comparing Fig. 4(b) with Fig. 3(b), there must be a transition 
with a critical set of $m$, $R$ and $M_g$ so that 
$J_{JY1}=J_{L2}$, i.e. 
$F(x_{JY1},y_{JY1},z_{JY1})=F(x_{L2},y_{L2},z_{L2})$.
In order to search the critical set of $m$, $R$ and $M_g$, 
we define a function :
\beqn
T_2(m,R,M_g)\non
&\equiv& F(x_{JY1},y_{JY1},z_{JY1})- F(x_{L2},y_{L2},z_{L2}) \non\\
&=& F(x_{JY1},0,0)-F(x_{L2},0,0) \non\\
&=& n^2(x^2_{JY1}-x^2_{L2}) +2m\left[\frac{1}{R-x_{JY1}}+\frac{1}{R+x_{JY1}}
-\frac{1}{x_{L2}-R}-\frac{1}{x_{L2}+R}\right] \non \\
&+&\frac{4M_g}{\pi}
\left(\frac{\tan^{-1}x_{JY1}}{x_{JY1}}-\frac{\tan^{-1}x_{L2}}{x_{L2}}\right).
\label{eq:T2}
\eeqn
A zero point of $T_2$ would correspond to a critical set of
$m$, $R$, and $M_g$.
When $T_2(m,R,M_g)=0$, we find that $M_g$ can be explicitly expressed as 
a function of $m, R$. We define this critical $M_g$ to be $M_2$:
\beq
M_2=\frac{\pi\left[n^2(x_{L2}^2-x_{JY1}^2)
+\frac{2m}{x_{L2}+R}+\frac{2m}{x_{L2}-R}-\frac{2m}{R-x_{JY1}}
-\frac{2m}{R+x_{JY1}}\right]}{4\left(\frac{\tan^{-1}x_{JY1}}{x_{JY1}}
-\frac{\tan^{-1}x_{L2}}{x_{L2}}\right)}.
\label{eq:m2}
\eeq
\no 

From Eq. (\ref{eq:m2}), we numerically solve
$M_2$  as a function of $R$ for different values of $m$, as shown in
Fig. 5(b). It is found that $M_2$ is a decreasing function of the radius 
$R$ for different values of $m$.
When $m=R=1$ and $M_g = 30$, 
the critical mass $M_2$ is huge, so $M_g < M_2$. 
When $m=1$, $R=2$, $M_g=30$, 
the critical mass $M_2$ is less than 20,
so $M_g > M_2$.
Therefore, the results in Figs.3 and 4 are 
on different sides of this transition.
 
\section{Mathematical Properties of 
the Jacobi Surfaces}

Mathematical properties of the Jacobi surfaces, 
i.e. $F(x,y,z)$, are investigated in this section.
From Eq.(\ref{eq:f1}), partial derivatives of the function $F(x,y,z)$
can be calculated as:
\beqn
&&\frac{\pa F}{\pa x}=2n^2x-\frac{2m(x+R)}{r_1^3}-\frac{2m(x-R)}{r_2^3}+
\frac{2cx}{r}\left\{\frac{1}{(1+r^2)r}-\frac{\tan^{-1}r}{r^2}\right\}
\label{eq:dcjdx} \\
&& \frac{\pa F}{\pa y}=2y\left\{n^2-\frac{m}{r_1^3}-\frac{m}{r_2^3}+
\frac{c}{r}\left[\frac{1}{(1+r^2)r}-\frac{\tan^{-1}r}{r^2}\right]\right\}
\label{eq:dcjdy} \\
&& \frac{\pa F}{\pa z}=2z\left\{-\frac{m}{r_1^3}-\frac{m}{r_2^3}+
\frac{c}{r}\left[\frac{1}{(1+r^2)r}-\frac{\tan^{-1}r}{r^2}\right]\right\}.
\label{eq:dcjdz}
\eeqn
From  Eq.(\ref{eq:dcjdx}), for convenience, we define :
\beq
\frac{\pa F}{\pa x}\biggm|_{(x,y,z)=(x,0,0)}\equiv A(x). \label {eq:ax}
\eeq
Compared with the equations in Jiang \& Yeh (2014a),
we realize that the zero points of
$A(x)$ are the equilibrium points on the $x$-axis.
Moreover, as shown in Jiang \& Yeh (2014a),
there is a critical mass $M_c$ which determines whether or not there
are Jiang-Yeh Points.
That is, there are JY1 and JY2 equilibrium points in the system
if, and only if, $M_g>M_c$.

\no{\bf Property 1:}\\ 
On the $x$-axis with $x_{L1}=0\le x \le x_{L2}$,\\
{\bf (a)} when $M_g<M_c$,
the maximum of $F(x,0,0)$ is
at the location of the black hole $(R,0,0)$,
and two relative minimum values of $F(x,0,0)$ are at
$0$ and $x_{L2}$; \\
{\bf (b)} when $M_g>M_c$, the maximum of $F(x,0,0)$
is at the location of the black hole $(R,0,0)$,
and two relative minimum values
of $F(x,0,0)$ are at $x_{JY1}$ and $x_{L2}$.

\no {\bf (Proof):}

According to Eq. (\ref{eq:dcjdx}), the equilibrium points of the system are
zero points of $A(x)$.  We first consider the function
$F(x,0,0)$ in the interval $(R, x_{L2})$,
then in the interval $(x_{L1},R)$ (where $x_{L1}$=0)
when $M_g<M_c$,
and finally in the interval $(x_{JY1},R)$ when $M_g>M_c$.

\no {\bf (1)} In the interval $(R, x_{L2})$:\\
Since there is one equilibrium point L2 at $(x_{L2},0,0)$ and no
equilibrium point in $(R, x_{L2})$, we have
$A(x_{L2})=0$ and $A(x)\ne 0$ for $R<x< x_{L2}$.
From Eq. (\ref{eq:dcjdx}), we find
$\lim_{x\to R^{+}}A(x)=-\infty$,  so
 $A(x)<0$ for $R<x< x_{L2}$. From the definition of
$A(x)$ in Eq. (\ref{eq:ax}), $F(x,0,0)$ is a decreasing function in
$(R, x_{L2})$.

\no {\bf (2)} In the interval $(x_{L1}, R)$ with $M_g<M_c$:\\
The equilibrium point L1 is at $(0,0,0)$, so $x_{L1}=0$ and $A(0)=0$. 
In this case with $M_g<M_c$, there is no equilibrium
point in $(0,R)$; thus, $A(x)\ne 0$ for $0<x< R$.  From Eq. (\ref{eq:dcjdx}),
we find $A(R^{-})=\infty$, which leads to
$A(x)> 0$ for $0<x<R$. Therefore, $F(x,0,0)$ is an increasing function
in $(0,R)$.

\no {\bf (3)} In the interval $(x_{JY1}, R)$ with $M_g>M_c$:\\
When $M_g>M_c$, there is an equilibrium point JY1 in $(0,R)$,
so $A(x_{JY1})=0$. There is no equilibrium
point in $(x_{JY1},R)$; thus, $A(x)\ne 0$ for $x_{JY1}<x<R$.
From Eq. (\ref{eq:dcjdx}), we find $A(R^{-})=\infty$,
which gives $A(x)>0 $ for $x_{JY1}<x<R$. That is,
$F(x,0,0)$ is an increasing function in $(x_{JY1},R)$.

The combination of {\bf (1)} and {\bf (2)} leads to that,
for the interval $(0, x_{L2})$ on $x$-axis, $F(x,0,0)$ has the maximum value
at $(R,0,0)$ and has local minimum values at $(0,0,0)$
and  $(x_{L2},0,0)$ when $M_g<M_c$. Thus, {\bf (a)} is proved.

The combination of {\bf (1)} and {\bf (3)} leads to that,
for the interval $(x_{JY1}, x_{L2})$ on $x$-axis,
$F(x,0,0)$ has the maximum value
at $(R,0,0)$ and has local minimum values at $(x_{JY1},0,0)$
and  $(x_{L2},0,0)$ when $M_g>M_c$. Thus, {\bf (b)} is proved.$\Box$\\

\no{\bf Property 2:}\\
{\bf (a)} Along the line $x=R$ of the $xy$-plane,
the maximum value of $F(R,y,0)$ is at the black hole $(R,0,0)$,
and there is a relative minimum of
$F(R,y,0)$ at a point $y^\ast>0$.  \\
{\bf (b)} Along the line $x=R$ of the $xz$-plane,
 the maximum value of $F(R,0,z)$ is at the black hole
 $(R,0,0)$.

\no {\bf (Proof):}\\
\no {\bf (a)} From Eq.(\ref{eq:dcjdy}), we define
\beqn
B_R(y)&=&\frac{\pa F}{\pa y}\biggm|_{(x,y,z)=(R,y,0)}=2y\left\{ n^2
-\frac{m}{r_1^3}-\frac{m}{r_2^3}+\frac{c}{r}\left[ \frac{1}{(1+r^2)r}
-\frac{\tan^{-1} r}{r^2}\right]\right\} \non \\
&\equiv&  2y\left[-P(y)+Q(y) \right],
\eeqn
\no where $P(y)=\frac{m}{r_1^3}+\frac{m}{r_2^3}-n^2$,
$Q(y)=\frac{c}{r}\left[\frac{1}{(1+r^2)r}-\frac{\tan^{-1}{r}}{r^2}\right]$,
$r_1^2=4R^2+y^2$, $r_2^2=y^2$, and $r^2=R^2+y^2$.
For positive $y$, since $\lim_{y\to 0} y P(y)=\infty$
and $\lim_{y\to 0}y Q(y)=0$, we know $\lim_{y\to 0}B_R(y)=-\infty$.
Moreover, because $\lim_{y\to \infty} y P(y)=-\infty$
and $\lim_{y\to \infty} yQ(y)=0$, we have
$\lim_{y\to \infty} B_R(y)=\infty$.
Thus, there is at least one point in the
interval $(0,\infty)$, such that $B_R(y)=0$. We assume
the smallest zero point to be $y^\ast \in (0,\infty)$
such that $B_R(y^\ast)=0$.
Since $B_R(y)=\frac{\pa F}{\pa y}|_{(R,y,0)}<0$ for $0<y<y^\ast$, there is
a relative minimum of $F(R,y,0)$ at point $y^\ast>0$.

Because $B_R(-y)= - B_R(y)$, 
we can similarly have a zero point $y^\ast_1 \in (-\infty,0)$  such that
$B_R(y^\ast_1)=0$. Since $B_R(y)=\frac{\pa F}{\pa y}|_{(R,y,0)}<0$ for
$0<y<y^\ast$ and  $B_R(y)=\frac{\pa F}{\pa y}|_{(R,y,0)}>0$
for $y^\ast_1<y<0$, there is a maximum value of
$F(R,y,0)$ at the black hole $(R,0,0)$ along the $y$-axis.

\no {\bf (b)} From Eqs.(\ref{eq:dcjdz}) and (\ref{eq:fg}), we have : 
$$\frac{\pa F}{\pa z}\biggm|_{(x,y,z)=(R,0,z)}=2z\left\{-\frac{m}{r_1^3}-\frac{m}{r_2^3}+\frac{c}{r}
\left[\frac{1}{(1+r^2)r}-\frac{\tan^{-1} r}{r^2}\right]\right\}=2z\left\{-\frac{m}{r_1^3}-\frac{m}{r_2^3}+\frac{f_g(r)}{r}\right\},$$ 

\no where $r_1^2=4R^2+z^2$, $r_2=|z|$, $r^2=R^2+z^2$.
Because  
the gravitational force $f_g(r)<0$ at any $r$,
$\frac{\pa F}{\pa z}<0$ for $z>0$ and
$\frac{\pa F}{\pa z}>0$ for $z<0$, there is a maximum value of
$F(R,0,z)$ at the black hole $(R,0,0)$ along the z-axis.$\Box$
\clearpage

\no{\bf Property 3:}\\
{\bf (a)} 
Along the $y$-axis of the $xy$-plane,
the maximum value of $F(0,y,0)$ is at the origin $(0,0,0)$. 
In addition, there is one relative minimum value of $F(0,y,0)$ at $y_{L4}$,
and another relative minimum value of $F(0,y,0)$ at $y_{L5}$.  \\
{\bf (b)} When $M_g>M_c$, we have $0<z_{d}<y_{d}<x_{JY1}$;
           $z_d$ and $y_d$ are defined in Section 3.\\
{\bf (c)} Along the $z$-axis of $xz$-plane,
the maximum value of $F(0,0,z)$ is at the origin $(0,0,0)$.

\no {\bf (Proof):}\\
\no {\bf (a)} From Eq.(\ref{eq:dcjdy}), for $y>0$, we define :
\beqn
B_0(y)&=&\frac{\pa F}{\pa y}\biggm|_{(x,y,z)=(0,y,0)}=2y\left\{ n^2
-\frac{m}{r_1^3}-\frac{m}{r_2^3}+\frac{c}{y}\left[ \frac{1}{(1+y^2)y}
-\frac{\tan^{-1} y}{y^2}\right]\right\} \non \\
&\equiv& 2y\left[-P(y)+Q(y) \right],
\eeqn
\no where $P(y)=\frac{m}{r_1^3}+\frac{m}{r_2^3}-n^2$,
$Q(y)=c\left[\frac{1}{(1+y^2)y^2}-\frac{\tan^{-1}{y}}{y^3}\right]$,
and $r_1^2=r_2^2=R^2+y^2$. 
From the above definition of $Q(y)$, we have :
\beq
\lim_{y\to 0} Q(y) = \lim_{y\to 0}\frac{c}{y^3}
\left(\frac{y}{1+y^2}-\tan^{-1}y\right)
=\lim_{y\to 0}\frac{-2cy^2}{3y^2(1+y^2)^2} =-\frac{2c}{3}<0,
\label{eq:q0}
\eeq 
and
\beq
Q'(y)= c\frac{-3y-5y^3+3(1+y^2)^2\tan^{-1}(y)}
{y^4(1+y^2)^2}
= c\left\{\frac{3(1+y^2)^2g_1(y)}{y^4(1+y^2)^2}\right\}, 
\label{eq:dq1}
\eeq
where $g_1(y)\equiv \tan^{-1}y-\frac{5y^3+3y}{3(1+y^2)^2}$.  Since 
$g_1'(y)=\frac{8y^4}{3(1+y^2)^3}>0$ for $y>0$ and $g_1(0)=0$, we 
have $g_1(y)>0$ for $y>0$. Thus, from Eq. (\ref{eq:dq1}), we know
$Q'(y)>0$ for all $y>0$. Therefore, from Eqs.(\ref{eq:q0}) 
and (\ref{eq:dq1}),
we have $-\frac{2c}{3}<Q(y)<0$ for $y>0$. Moreover, 
from Eqs.(\ref{eq:q0}) and (\ref{eq:n2}), we have :

\beq
\lim_{y\to 0} \left[-P(y)+Q(y)\right]=-\frac{2m}{R^3}+n^2-\frac{2c}{3}
= -\frac{7m}{4R^3}+\left[|Q(R)|-\frac{2c}{3}\right]<0, \label{eq:pq0}
\eeq

\no so $\lim_{y\to 0}B_0(y)= \lim_{y\to 0} 2y[-P(y)+Q(y)]=0$. 
Since $B'_0(y)=2[-P(y)+Q(y)]+2y[-P'(y)+Q'(y)]$, from Eq.(\ref{eq:pq0}),
 we have 
$B'_0(0)=2[-P(0)+Q(0)]<0. $  According to Eq.(\ref{eq:dcjdy}), the 
equilibrium points of the system are zero points of $B_0(y)$. Since there 
is one equilibrium point L4 at $(0,y_{L4},0)$ and no equilibrium point in 
$y\in (0,y_{L4})$, we have $B_0(0)=B_0(y_{L4})=0$, $B_0(y)\ne 0$ for 
$0<y<y_{L4}$ and $ B'_0 (0)<0$. Therefore, 
$B_0(y)=\frac{\pa F}{\pa y}|_{(x,y,z)=(0,y,0)}<0$ for $0<y<y_{L4}$;
there is a relative minimum value of $F(0,y,0)$ at point $y_{L4}$.

Through a similar method, it can be shown that 
$\frac{\pa F}{\pa y}|_{(x,y,z)=(0,y,0)}>0$ for $y_{L5}<y<0$,
and there is a relative minimum value of $F(0,y,0)$ at point $y_{L5}$.
Therefore, along the $y$-axis of $xy$-plane,
the maximum value of $F(0,y,0)$ is at the origin $(0,0,0)$.

\no {\bf (b)} When $M_g>M_c$, JY1 exists.  
 As $F(x_{JY1},0,0)$ is the value of Jacobi integral at JY1, 
we set $F(x_{JY1},0,0)=J_{JY1}$.
Thus, $F(x,y,0)=J_{JY1}$ is the JY1-passing equi-potential curve 
on $xy$-plane.
Since $(0,y_d,0)$ is the intersection between $F(x,y,0)=J_{JY1}$ and $y$-axis,
$y_d$ is the root of $F(0,y,0)=J_{JY1}$.
To examine the existence of $y_d$, 
we define :
\beqn
&&k_y(y)\equiv  F(0,y,0)-F(x_{JY1},0,0) =n^2(y^2-x_{JY1}^2) \non \\
&& +2m\left(\frac{2}{\sqrt{R^2+y^2}}-\frac{1}{R+x_{JY1}}
-\frac{1}{R-x_{JY1}}\right)+2c\left[\frac{\tan^{-1}|y|}{|y|}-
\frac{\tan^{-1}x_{JY1}}{x_{JY1}}\right].\label{eq:ky}
\eeqn

From Fig. \ref{fig:ky_0}, we find that $k_y(0)>0$ 
for considered values of $m$, $R$, and $M_g$. 
From Eq.(\ref{eq:ky}), we have :
\beqn
&&k_y(x_{JY1})=2m\left(\frac{2}{\sqrt{R^2+x^2_{JY1}}}-\frac{1}{R+x_{JY1}}
-\frac{1}{R-x_{JY1}}\right) \non \\
&&
=2m\left(\frac{2(R^2-x_{JY1}^2)-2R\sqrt{R^2+x_{JY1}^2}}{\sqrt{R^2+x_{JY1}^2}
(R+x_{JY1})(R-x_{JY1})}\right)=4m\left(
\frac{R(R-\sqrt{R^2+x_{JY1}^2})-x_{JY1}^2}
{\sqrt{R^2+x_{JY1}^2}(R+x_{JY1})(R-x_{JY1})}\right)<0. \non
\eeqn
Thus, there is a point $y_{d}\in (0,x_{JY1})$ such that $k_y(y_{d})=0$.

Further, we define :
\beqn
&&k_z(z)= F(0,0,z)-F(x_{JY1},0,0)=-n^2x_{JY1}^2 \non \\
 && +2m\left(\frac{2} {\sqrt{R^2+z^2}}-\frac{1}{R+x_{JY1}}
-\frac{1}{R-x_{JY1}}\right)+2c\left[\frac{\tan^{-1}|z|}{|z|}-
\frac{\tan^{-1}x_{JY1}}{x_{JY1}}\right].\label{eq:kz}
\eeqn

\no  From Eqs.(\ref{eq:ky})-(\ref{eq:kz}), we
have $k_z(0)=k_y(0)>0$. 
In addition, we also have :
\beqn
 k_z(y_d) &=& -n^2x_{JY1}^2
+2m\left(\frac{2}{\sqrt{R^2+y_{d}^2}}-\frac{1}{R+x_{JY1}}
-\frac{1}{R-x_{JY1}}\right)+2c\left[\frac{\tan^{-1}y_{d}}{y_{d}}-
\frac{\tan^{-1}x_{JY1}}{x_{JY1}}\right] \non \\
&=& k_y(y_{d}) - n^2y_{d}^2. \non
\eeqn

\no Because $k_y(y_{d})=0$,  we have $k_z(y_{d})=-n^2y_{d}^2<0$. Thus, 
there is a point $z_{d}\in (0,y_{d})$ such that $k_z(z_{d})=0$. Therefore, 
we have $0<z_{d}<y_{d}<x_{JY1}$.

\no {\bf (c)}  From Eqs.(\ref{eq:dcjdz}) and (\ref{eq:fg}), we have 
$$\frac{\pa F}{\pa z}\biggm|_{(x,y,z)=(0,0,z)}=2z\left\{-\frac{m}{r_1^3}
-\frac{m}{r_2^3}+\frac{c}{r}\left[\frac{1}{(1+r^2)r}
-\frac{\tan^{-1} r}{r^2}\right]\right\}=2z\left\{-\frac{m}{r_1^3}
-\frac{m}{r_2^3}+\frac{f_g(r)}{r}\right\},$$ 

\no where $r_1^2 = r_2^2 = R^2+z^2 $, $r=|z|$. Since
the gravitational force $f_g(r)<0$, 
we have $\frac{\pa F}{\pa z}<0$ for $z>0$ and
$\frac{\pa F}{\pa z}>0$ for $z<0$. Therefore, along the $z$-axis, there is 
a maximum value of $F(0,0,z)$ at the origin $(0,0,0)$.$\Box$

\section{The Determination of Lobe Regions}

In order to know whether a particle is inside
the Roche Lobes or Jiang-Yeh Lobe, we need a standard procedure
to determine the regions of lobes.
One possibility is to numerically determine
the Roche Lobes or Jiang-Yeh Lobe and keep the coordinates
of these surfaces. To judge whether a particle is inside these
closed surfaces, we can simply compare the particle's coordinate with
the coordinate of this particle's corresponding point projected 
onto these surface along one of the axes
of coordinate system. However, one would need to do
a multi-dimensional numerical interpretation
to get this projection done. When particles are very close to
one of these surfaces, one would need much higher resolution on the numerical
coordinates of surfaces, so it would be a very time-consuming
process.

In order to avoid this difficult, we propose another 
procedure which is based on
the monotonic behavior of the Jacobi integral of equi-potential surfaces.
After the coordinates of equilibrium points and other important points 
are found, the smallest radius, $r_R$, of the shell 
which can enclose the Roche Lobe
centered on $(R,0,0)$ can be determined.
Similarly, the smallest 
radius, $r_{JY}$, of the shell enclosing the Jiang-Yeh Lobe
centered on $(0,0,0)$ can be determined, too.
For example, in Model A,
$r_R$ is set as the largest among
$|x_a-R|$, $|R-x_{L1}|$, $y_a$, $z_a$, 
so it is the smallest radius which
can enclose the Roche Lobe.
Similarly, for Model B, $r_R$ is the largest among
$|x_b-R|, |R-x_{L2}|, y_b, z_b$.
For Model C, the smallest radius $r_{R}$ that can enclose
Roche Lobe is defined in the same way as in Model B; and
the smallest radius $r_{JY}$ that can enclose
Jiang-Yeh Lobe is defined as
the largest among
$x_{JY1}$, $y_d$ and $z_d$.
For Model D, the smallest radius $r_{R}$ that can enclose
Roche Lobe is the largest among
$|x_c-R|, |R-x_{JY1}|, y_c, z_c$; and
the smallest radius $r_{JY}$ that can enclose
Jiang-Yeh Lobe is determined in the same way as in Model C.
After $r_R$ and $r_{JY}$ are determined, we then need to confirm that 
the Jacobi integral of equi-potential surfaces is either 
monotonically decreasing or monotonically 
increasing from the centers of the lobes. Through the comparisons 
between the values of Jacobi integrals of equi-potential surfaces
at a particle's coordinate 
and at the boundary of a particular lobe, we can know  whether this particle 
is in the lobe region.   
The exact steps are listed as follows:\\

\no {\bf Step 1.} Determine the locations of equilibrium points numerically
i.e. $(x_{L1}, y_{L1},0)$,
      $(x_{L2}, y_{L2},0)$,
      $(x_{L3}, y_{L3},0)$,
      $(x_{L4}, y_{L4},0)$,
      $(x_{L5}, y_{L5},0)$,
      $(x_{JY1}, y_{JY1},0)$,
      $(x_{JY2}, y_{JY2},0)$.
Please note that
$x_{L1}=y_{L1}=y_{L2}=y_{L3}=x_{L4}=x_{L5}=y_{JY1}=y_{JY2}=0$.

\no {\bf Step 2.} Draw zero-velocity curves which pass
      through L1, L2 and JY1 on $xy$-plane and $yz$-plane.
  Then, numerically determine the intersection points 
  between these curves and coordinate axes,
   and also the intersection points between these curves
  and straight lines which go through the lobe center and
      are parallel with $x$-axis, $y$-axis, 
      or $z$-axis. For example, those marked points $x_a, x_b, y_b$... etc. 
      in Figs. 1-4.
       
\no {\bf Step 3.} Determine the smallest radius of shells which can enclose
the lobes, i.e. $r_R$ and $r_{JY}$.

\no {\bf Step 4.} Numerically show that $F(x,y,z)$
is monotonically decreasing or increasing 
from the center to the boundaries of these shells.
If
$\frac{\pa F}{\pa x}<0$, 
$\frac{\pa F}{\pa y}<0$, and
$\frac{\pa F}{\pa z}<0$
inside shells,  
$F(x,y,z)$ is monotonically decreasing
from the centers to the boundaries.
When
$\frac{\pa F}{\pa x}>0$, 
$\frac{\pa F}{\pa y}>0$, and
$\frac{\pa F}{\pa z}>0$
inside shells, then 
$F(x,y,z)$ is monotonically increasing
from the centers to the boundaries.

\no {\bf Step 5.} Compare the values of $F(x,y,z)$.
For a particle with coordinates $(x_p,y_p,z_p)$, 
we first determine whether it is inside one of the shells that enclose
Roche Lobes or Jiang-Yeh Lobe.
If it is inside the shell enclosing one of the Roche Lobes,
we use the value $F(x_p,y_p,z_p)$ to judge whether it is
inside or outside of the Roche Lobe by comparing its value with
$F(x_{L1},y_{L1},0)$ when two Roche Lobes are connected 
(as the case in Model A),
and by comparing its value with
$F(x_{L2},y_{L2},0)$ when two Roche Lobes are separated 
(as the case in Model B).
If it is inside the shell enclosing Jiang-Yeh Lobe,
we use the value $F(x_p,y_p,z_p)$ to judge whether it is
inside or outside of the Jiang-Yeh Lobe by comparing its value with
$F(x_{JY1},y_{JY1},0)$.

Therefore, through the above five steps, whether a particle is inside 
one of lobe regions can be determined completely 
by numerical calculations. 
In the case that some mathematical properties of the system are studied 
analytically, the known analytical results, as those presented 
in previous section, can be used to reconfirm parts of numerical 
calculations in Steps 3 and 4.
The numerical values of coordinates of those important points in Model A-D are 
presented in Table 1. These values determined by numerical calculations 
completely satisfy the mathematical properties we have proven.

\section{Concluding Remarks}

We have studied a galactic model with supermassive binary holes.
Assuming two black holes have equal mass,
the three-dimensional equi-potential surfaces, 
i.e. Roche Lobes and Jiang-Yeh Lobe are presented.
The critical conditions of two kinds of topological transitions
for these equi-potential surfaces were
investigated herein. When the galactic total mass $M_g=0$, two Roche 
Lobes are connected at the first Lagrange Point. 
When $M_g$ is larger, in one case, two Roche Lobes would separate first,
and a detached Jiang-Yeh Lobe shows up in the middle for 
an even larger value of $M_g$; in another case, the Jiang-Yeh Lobe appears 
directly and three lobes are connected.  
The mathematical properties of the 
Jacobi surfaces,  
$F(x,y,z)$, are also investigated.
In addition, a numerical procedure to determine the regions of 
Roche Lobes and Jiang-Yeh Lobe is described herein.
 
The main limitation of our model is that two supermassive black holes
are assumed to have equal mass and move in a circular orbit.
Thus, our model might not be used to study the galactic minor mergers,
but could be a good approximation to investigate the process
of galactic major mergers.
In a galactic major merger, both involved systems would 
have similar configurations.
Due to the dynamical friction, both supermassive black holes in galaxies
migrate into 
the central region until most surrounding stars are scattered out 
and the dynamical friction becomes very small. These two supermassive 
black holes will form an equal mass binary system and move in a circular orbit.
This is the stage during the galactic evolution that our model can act 
as a good tool to investigate 
the distribution and kinematics of stars in the system. 
Therefore, our model can lead to a prediction of
surface brightness and line-of-sight velocity distribution
of the galactic systems which expereinced major mergers. 
Future observations can test these predictions 
and the roles of 
Jiang-Yeh Lobe and Roche Lobes will be clear after comparisions 
between theoretical predictions and observational results 
are performed.

\section*{Acknowledgment}
We are grateful to the referee for good suggestions.
This work is supported in part 
by the Ministry of Science and Technology, Taiwan, under 
Li-Chin Yeh's 
Grants MOST 104-2115-M-134-004
and Ing-Guey Jiang's
Grants MOST 103-2112-M-007-020-MY3.

\clearpage
\vskip 0.1truein
\center{Table 1}
\begin{table}[ht]
\begin{center}
\begin{tabular}{ccccc}
\hline
 Model     &  A  &  B  &  C  &  D \\
\hline
$M_g$     &  0  & 10     & 30   & 30         \\
$m$       &  1  &  1    &  1  &   1      \\
$R$       &  1  &  1    &  1  &   2      \\
 $x_{L1}$ & 0.0   &  0.0    & 0.0   &     0.0    \\
 $x_{L2}$ & 2.3968 & 1.6989 &   1.5014       &  2.6753      \\
$x_{JY1}$ &  --  &   --     &  0.3608  &   1.3646  \\
$x_a$     & 1.810   &  --    &  --  &  --       \\
$x_b$     & --   &   0.4396  & 0.5932   & --       \\
$x_c$     & --   &  --    &--    &   2.5426     \\
$y_{L4}$  & 1.7320    &   1.1517     &  1.0604   &   2.0318      \\
$y_a$     & 0.7481     & 0.5596     & --   & --        \\
$y_b$     & 0.9646  & 0.4727     & 0.3380   & 0.4463        \\
$y_c$     & --   & --   & 0.3813   &   0.4275      \\
$y_d$     & --   & --   & 0.1498   &  1.039      \\
$z_a$     & 0.7123   & 0.4881     &  --  &  --       \\
$z_b$     & 0.8732   &  0.4320     &  0.3091   &  0.4180      \\
$z_c$     & --   &  --     & 0.3382   &  0.4033      \\
$z_d$     & --   &  --    & 0.1165   &  0.8751      \\
$r_R$     & 1   & 0.6989  & 0.5014   & 1.5426      \\
$r_{JY}$  & --   & --     & 0.3608   & 1.3646        \\
\hline
\end{tabular}
\caption[Table 1]{
The values of parameters, 
coordinates, and radii in Model A,B,C,D.  
}
\end{center}
\end{table}

\clearpage
\begin{figure}[ht]
\centering
\includegraphics[width=1\textwidth]{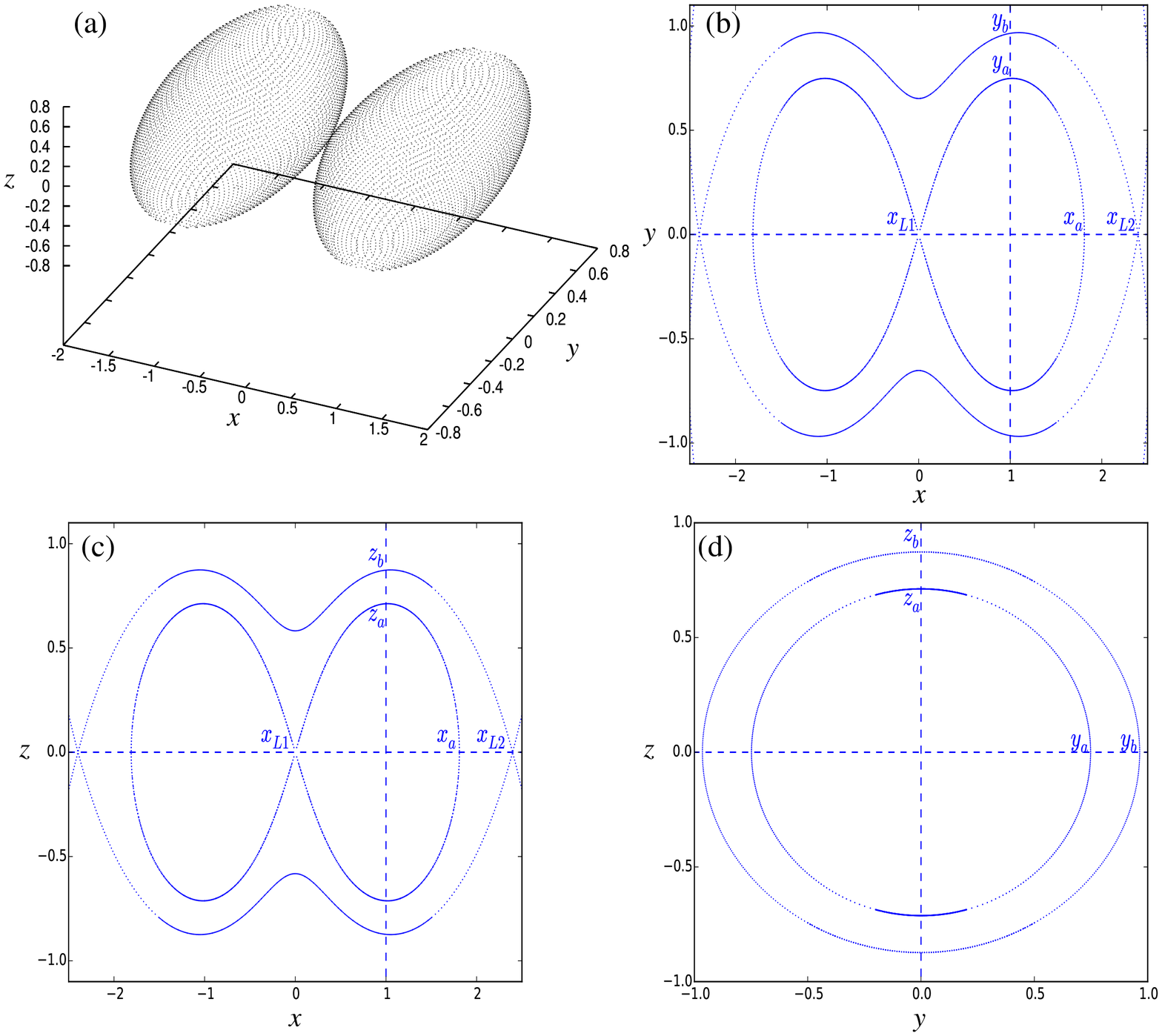}
\caption{The Roche Lobes and equi-potential curves 
when $m=R=1$ and $M_g=0$ (Model A). 
(a)The Roche Lobes in three dimensional space.
(b)The equi-potential curves on the $xy$-plane with $z=0$.
(c)The equi-potential curves on the $xz$-plane with $y=0$. 
(d)The equi-potential curves on the $yz$-plane with $x=R$.
}
\end{figure}

\clearpage
\begin{figure}[ht]
\centering
\includegraphics[width=1\textwidth]{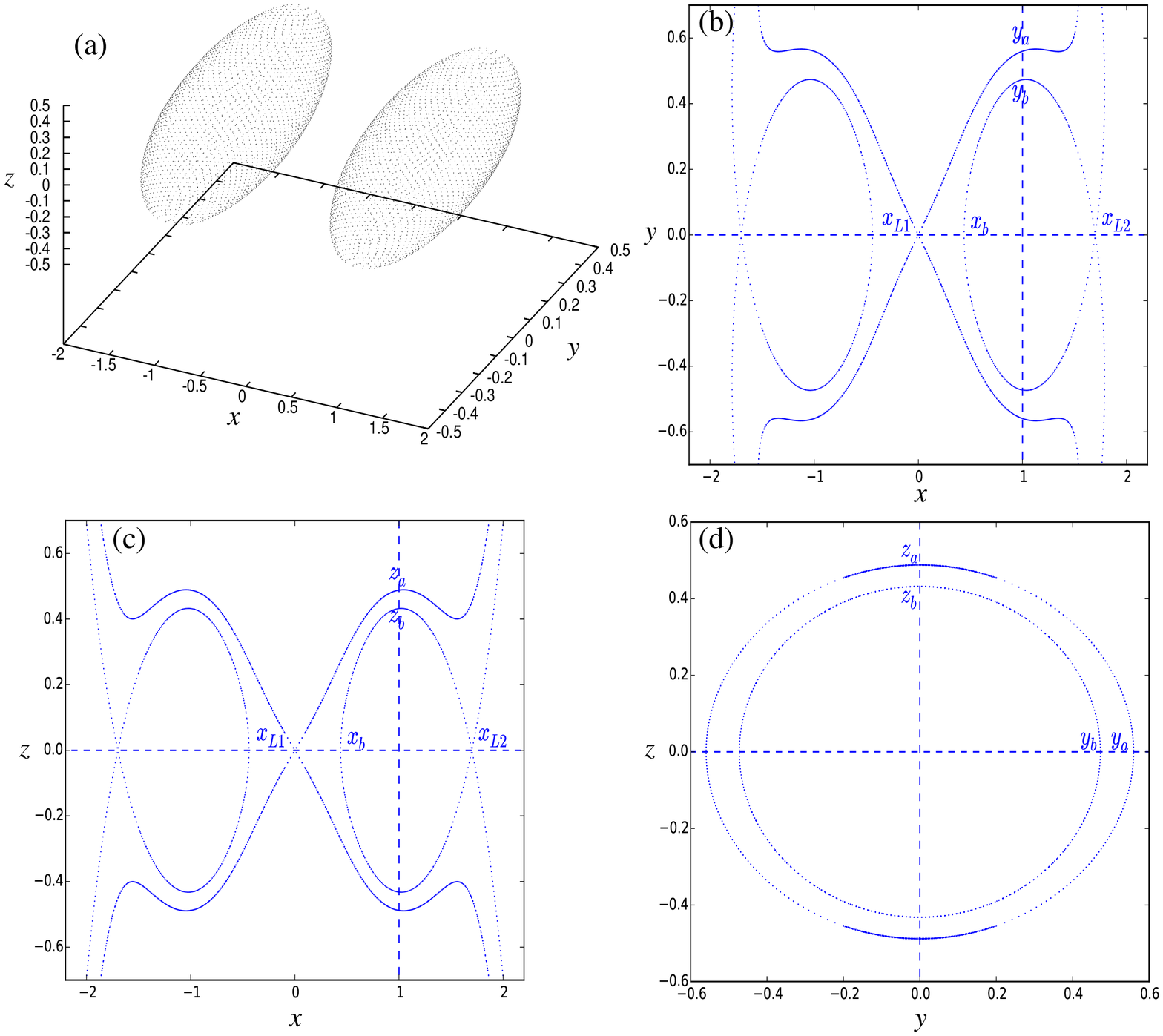}
\caption{The Roche Lobes and equi-potential curves 
when $m=R=1$ and $M_g=10$ (Model B). 
(a)The Roche Lobes in three dimensional space.
(b)The equi-potential curves on the $xy$-plane with $z=0$.
(c)The equi-potential curves on the $xz$-plane with $y=0$. 
(d)The equi-potential curves on the $yz$-plane with $x=R$.
}
\end{figure}  

\clearpage
\begin{figure}[ht]
\centering
\includegraphics[width=1\textwidth]{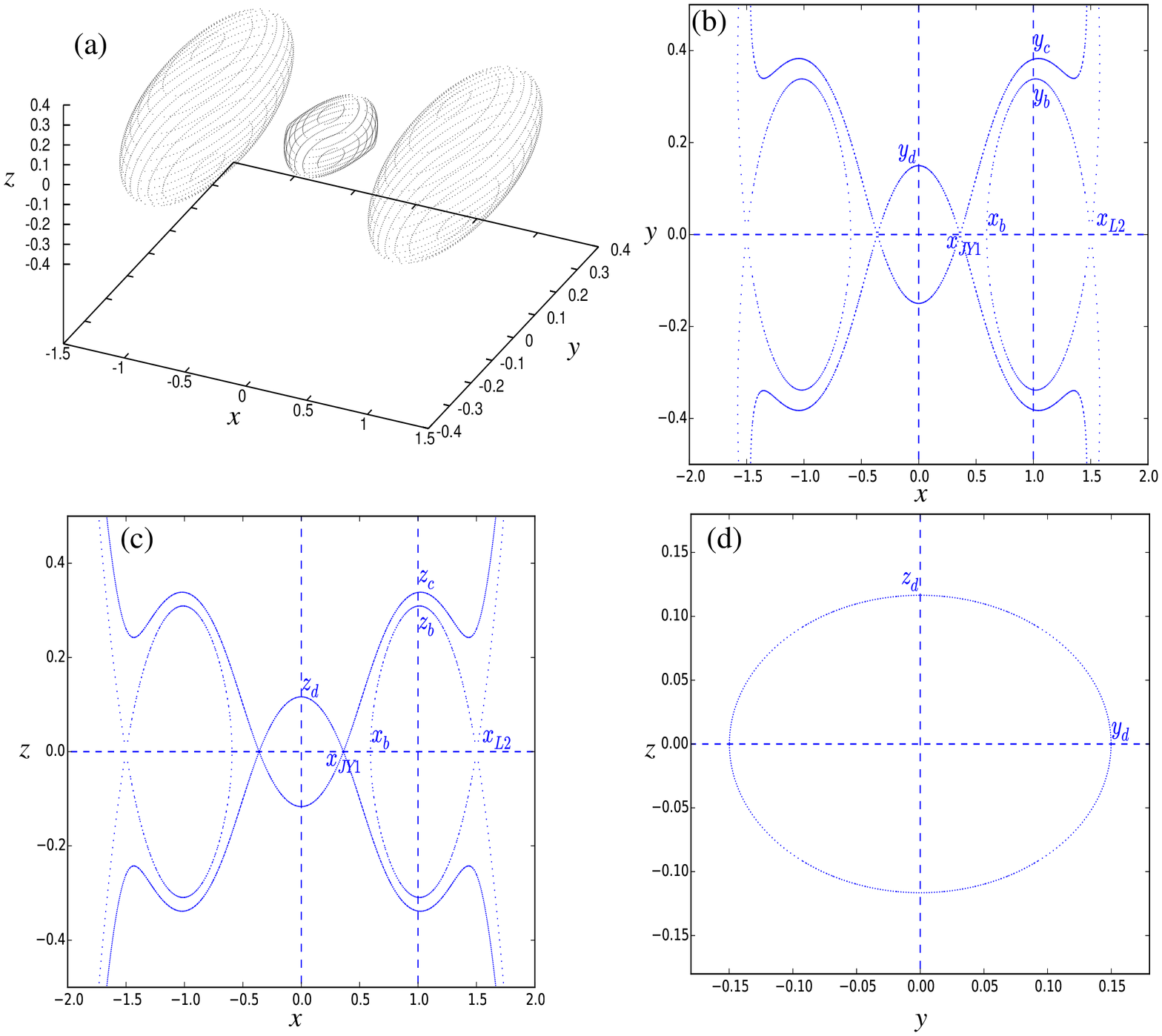}
\caption{The Roche Lobes, Jiang-Yeh Lobe, 
and equi-potential curves when $m=R=1$ and $M_g=30$ (Model C). 
(a)The Roche Lobes and Jiang-Yeh Lobe in three dimensional space.
(b)The equi-potential curves on the $xy$-plane $z=0$.
(c)The equi-potential curves on the $xz$-plane $y=0$. 
(d)The equi-potential curves on the $yz$-plane with $x=0$.
} 
\label{fig:cj_m30}
\end{figure} 

\clearpage
\begin{figure}[ht]
\centering
\includegraphics[width=1\textwidth]{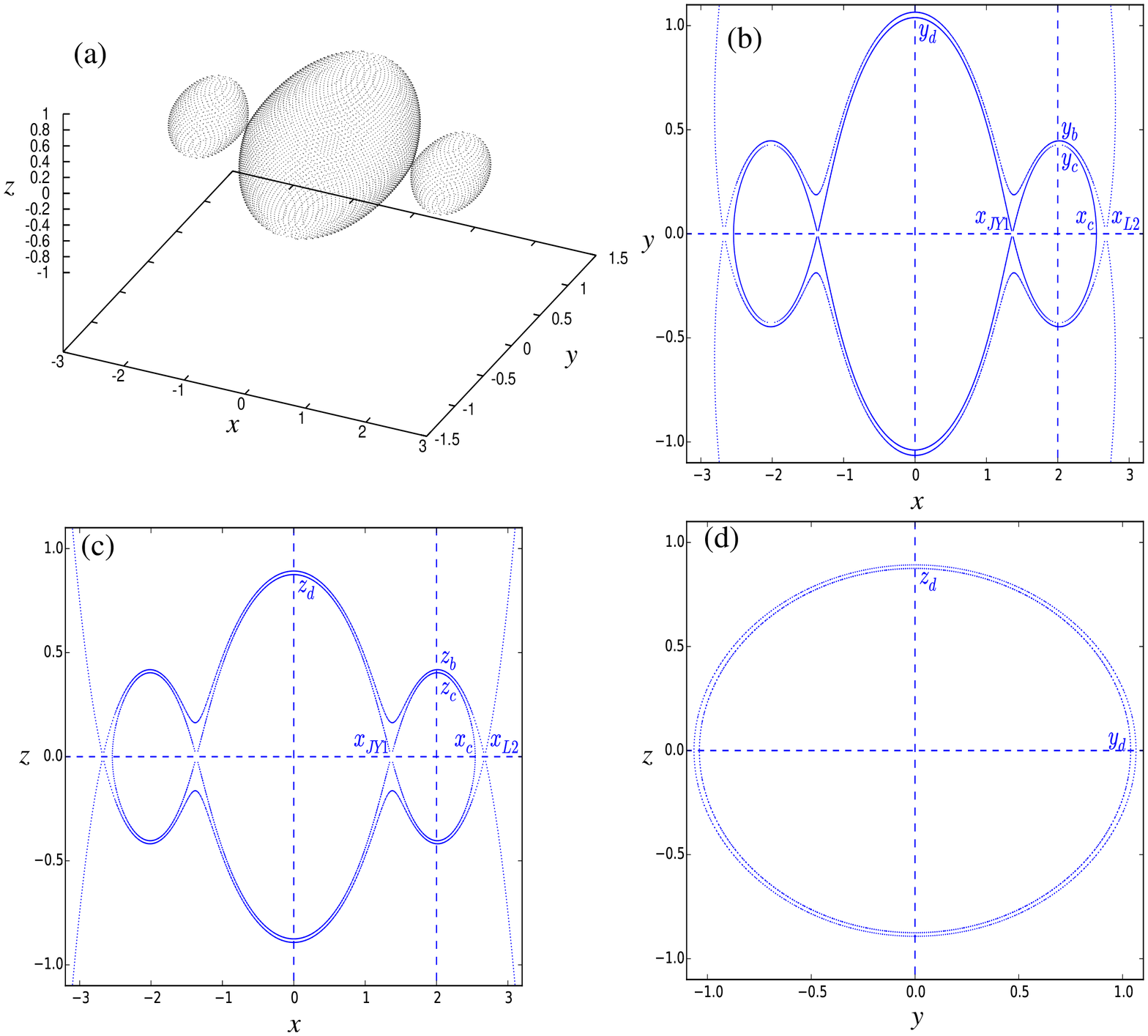}
\caption{The Roche Lobes, Jiang-Yeh Lobe, and equi-potential curves 
when $m=1$, $R=2$, and $M_g=30$ (Model D). 
(a)The Roche Lobes and Jiang-Yeh Lobe in three dimensional space.
(b)The equi-potential curves on the $xy$-plane $z=0$.
(c)The equi-potential curves on the $xz$-plane $y=0$. 
(d)The equi-potential curves on the $yz$-plane with $x=0$.
}
\label{fig:cj_m1r2}
\end{figure}  

\clearpage
\begin{figure}[ht]
\centering
\includegraphics[width=1\textwidth]{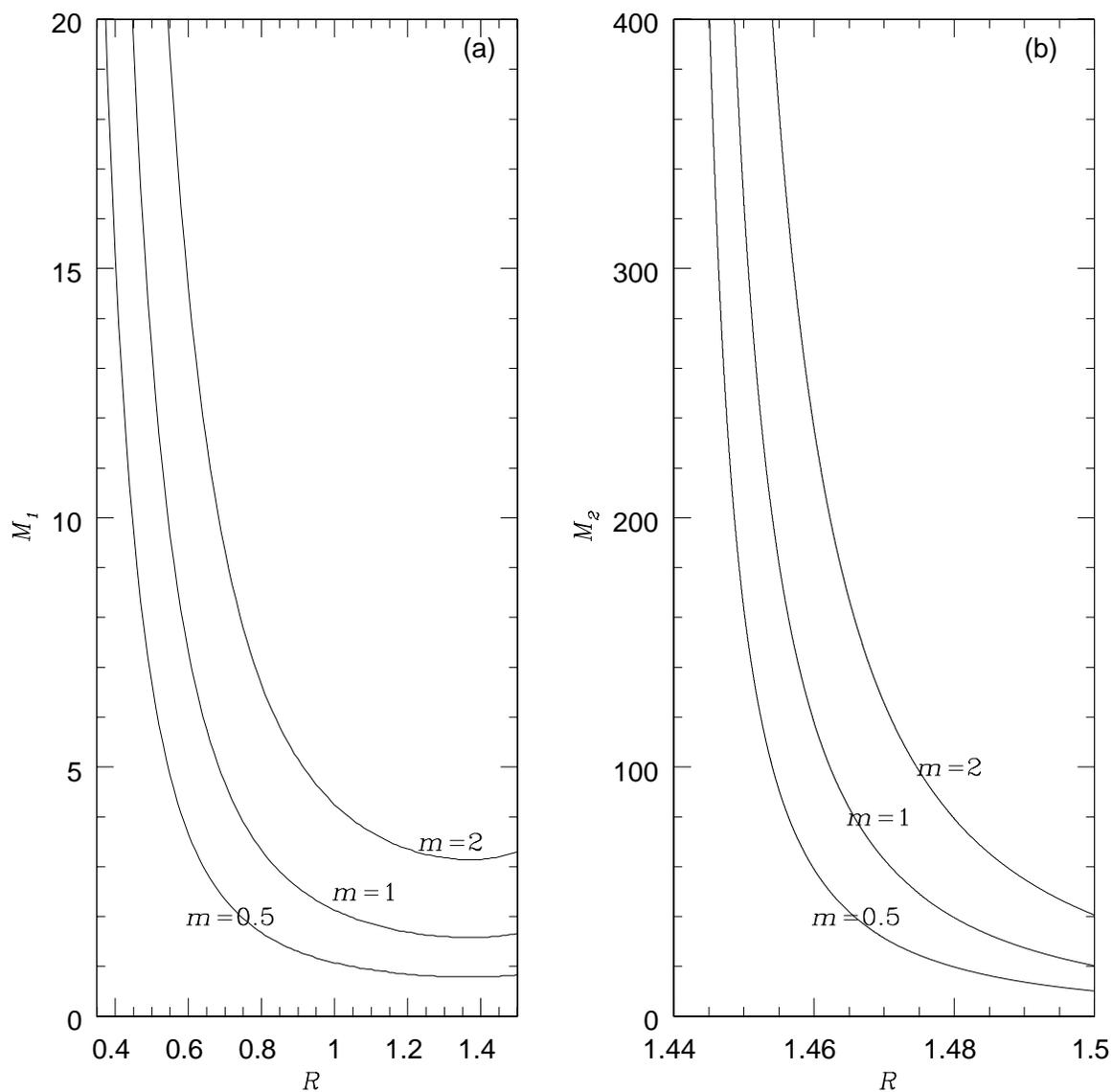}
\caption{The critical masses of transitions. 
(a) $M_1$ as a function of $R$ for different values of $m$. 
(b) $M_2$ as a function of $R$ for different values of $m$.
}
\label{fig:mc}
\end{figure}  

\clearpage
\begin{figure}[ht]
\centering
\includegraphics[width=1\textwidth]{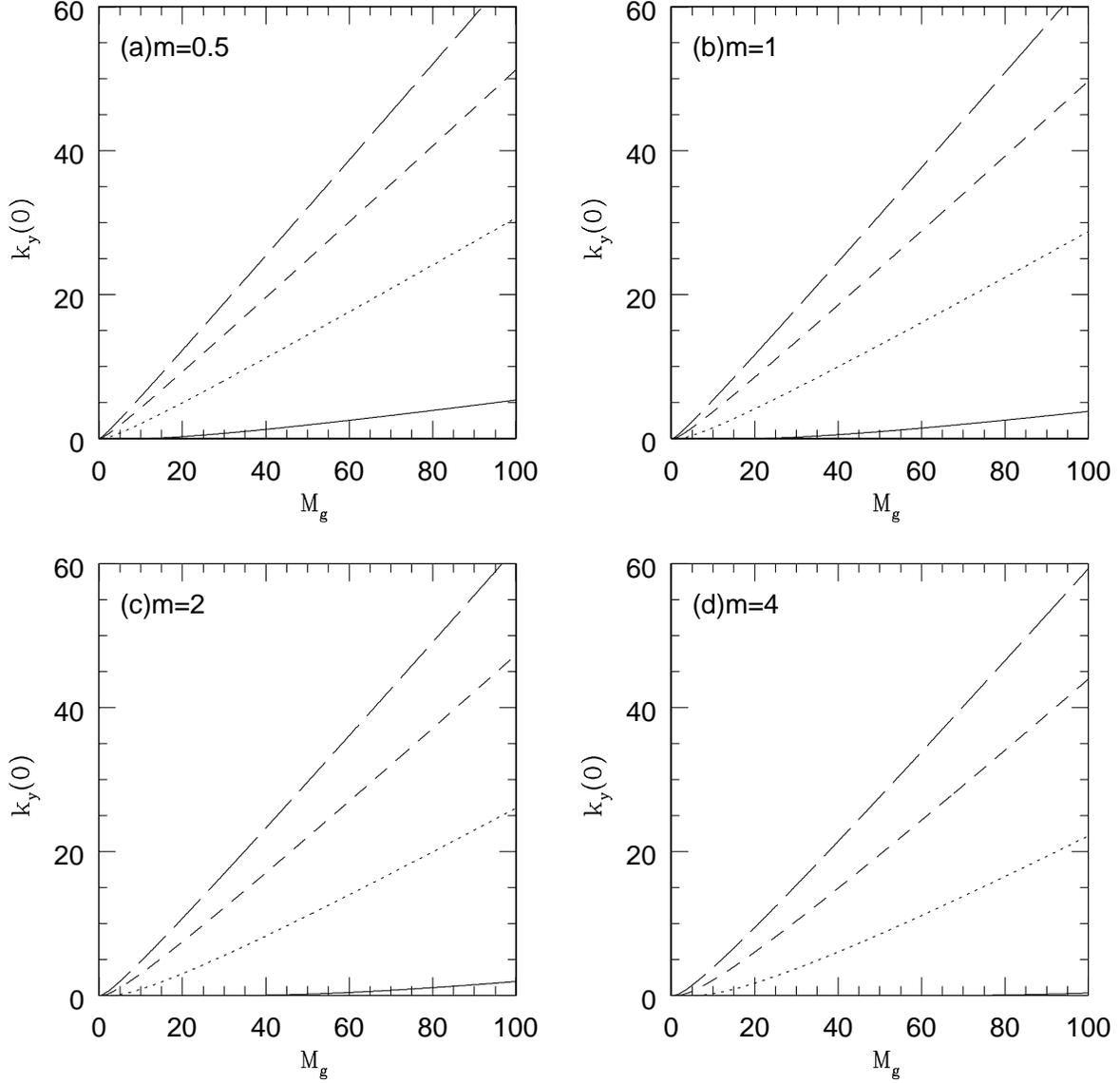}
\caption{The values of $k_{y}(0)$ as a function of $M_g$ for different 
values of $m$ and $R$. (a) is for $m=0.5$; (b) is for $m=1$; 
(c) is for $m=2$; (d) is for $m=4$.
In all panels, the solid line is for $R=1$;
the dotted line is for $R=2$; the dashed line is for $R=3$; and 
the long-dashed line is for $R=4$.} \label{fig:ky_0}
\end{figure}

\end{document}